\documentclass{moriond}
\usepackage{pstricks,pst-node,pst-text,pst-3d}
\usepackage{lineno}
\usepackage{booktabs}
\usepackage{amsmath}
\usepackage{placeins}
\usepackage{amssymb}
\usepackage{threeparttable}
\usepackage{graphicx}
\usepackage{lscape}
\usepackage[normalem]{ulem}
\usepackage[T1]{fontenc}
\usepackage[utf8]{inputenc}
\usepackage{setspace}
\usepackage{cite}
\usepackage{verbatim}
\usepackage{amsmath}
\usepackage{cleveref}

\usepackage{hyperref}
\usepackage{xcolor}
\hypersetup{
  colorlinks   = true, 
  urlcolor     = red, 
  linkcolor    = blue, 
  citecolor   = blue 
}

\crefname{figure}{Figure}{Figures}
\Crefname{figure}{Figure}{Figures}

\newcommand{\Photo}{\includegraphics[height=1mm]{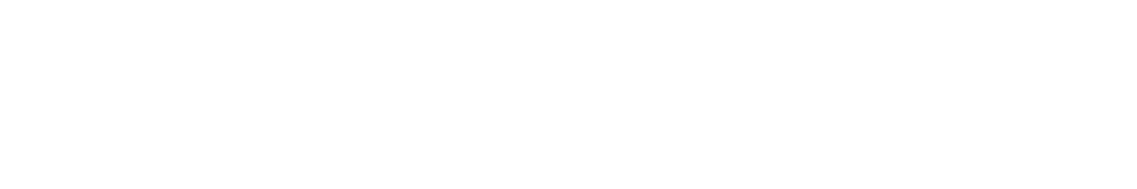}}

\begin{document}
\title{Dark Matter Searches with LUX}

\author{Cláudio Frederico Pascoal da Silva\\(for the LUX Collaboration)} 


\address{LIP-Coimbra \& Universidade de Coimbra, Rua Larga, Coimbra, Portugal}


\maketitle 

\abstract{We report here on the results from the Weakly Interacting Massive Particle (WIMP) search with the LUX dark matter experiment. LUX, a two-phase xenon time projection chamber (TPC) with 250 kg of active mass, has been operated from 2012 until 2016. During the operation, we observed no evidence for WIMP elastic scattering events. LUX achieved the most stringent limit on both the WIMP-nucleon spin-independent cross section (1.1$\times{10}^{-46}$~cm$^2$ for a 50 GeV/c${}^{2}$, 90\% C.L.) and on the WIMP-neutron spin-dependent cross section (1.6$\times{10}^{-41}$~cm$^2$ for a 35 GeV/c${}^{2}$, 90\% C.L.).}

\section{Introduction}
 The Large Underground Xenon (LUX)  experiment main objective is to study the nature of dark matter by detecting (or excluding) elastic scattering interactions of WIMPs\cite{Bertone2005}. The existence of cold dark matter is  supported by several cosmological observations such as galaxy dynamics\cite{Salucci2007}, cosmic microwave background\cite{Planck2015},  galaxy clusters, gravitational lensing, etc. The combination of these astrophysical observations result in a cold dark matter density of $\Omega_c$=0.258$\pm$0.011, significantly larger than the density of ordinary baryonic matter $\Omega_b$=0.0484$\pm$0.0010\cite{Planck2015}\footnote{$\Omega_c=\rho_c/\rho_{\mathrm{crit.}}$ and $\Omega_b=\rho_b/\rho_{\mathrm{crit.}}$, where $\rho_c$ and $\rho_b$ are the densities of cold dark matter and baryonic matter respectively, and $\rho_{\mathrm{crit.}}$ is the critical density of the Friedmann (flat) Universe.}. Several dark matter candidates have been proposed such as primordial black holes, axions, Weakly Interacting Massive Particles\cite{Goodman1985} (WIMPs) etc. From all of them, WIMPs are considered as the most promising candidates that will be discussed here.
 
 Direct detection experiments like LUX aim to detect nuclear recoils produced in the elastic scattering of a dark matter particle with the target nucleus. The observation of such recoils is challenging due to the very low interaction cross-section expected for the dark matter particle, compared with the much higher event rate from backgrounds (from detector materials and environment). Therefore, large detectors with very low background and low energy threshold are paramount for the direct WIMP detection.
 
LUX was the most sensitive direct detection experiment from 2013 until the recent results from XENON-1T\cite{XENON1T2017}. The detector operated inside the Davis Cavern, at the Sanford Underground Research Facility (SURF), USA,  at a depth of about 1480~m, from 2012 until May of 2016. The LUX detector had two different science runs, called here WS2013 and WS2014--16. WS2013 data were collected from April until August 2013 (95 live-days) and WS2014--16 were collected from September 2014 until May 2016 (total of 332~live-days). The analysis of the WS2013 data\cite{LUX2014_OriginalResults} (limited to 85.3~live-days) set a 90\% C.L.\ upper limit on the WIMP-nucleon spin-independent cross section of 0.76$\times$10$^{-45}$~cm$^2$ at a WIMP mass of 33 GeV/c$^2$. The re-analysis of the extended WS2013 dataset (total of 95~live-days) improved this limit to 0.60$\times$10$^{-45}$~cm$^2$ at 33~GeV/c$^2$ WIMP mass. Spin-dependent results\cite{LUXSD2016} for the WS2013 data at a WIMP mass of 33~GeV/c$^2$ set an upper limit of $\sigma_n$=9.1$\times$10$^{-41}$~cm$^2$ for the WIMP-neutron cross section and $\sigma_p$=2.9$\times$10$^{-39}$~cm$^2$ for a WIMP-proton cross section (90\% C.L.). Here, we report the results for both  the spin-dependent and spin-independent cross section for the combination of WS2013 and WS2014--16 datasets. These results have been first published in \cite{LUX2016_SSR} and \cite{LUX2017_SDCombined}. 

\section{The LUX detector}
\begin{figure}
\begin{center}
\includegraphics[width=8.5cm]{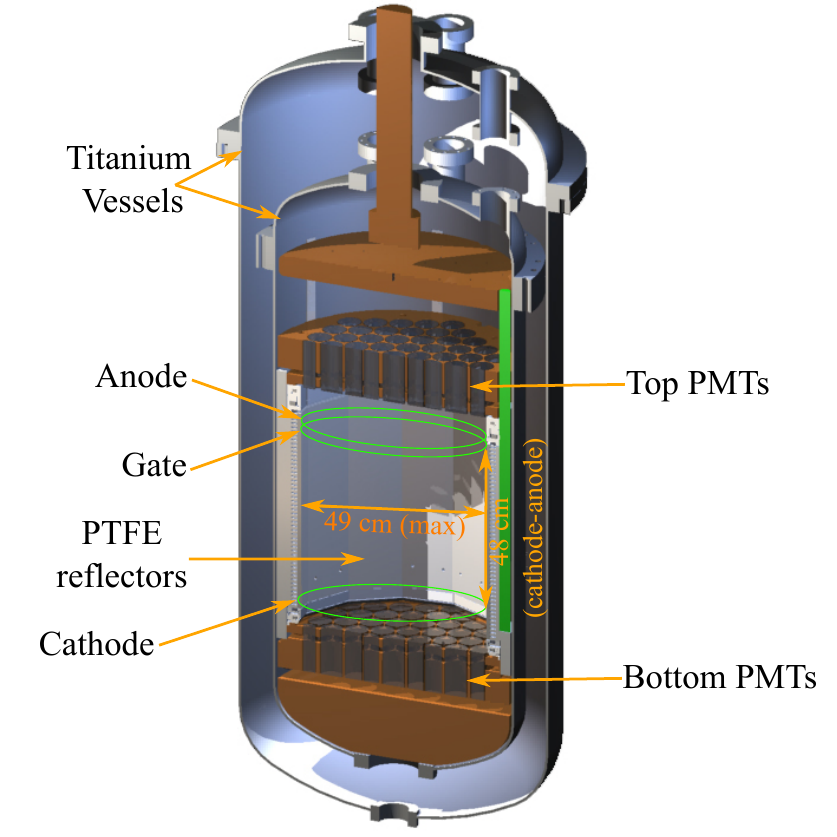}
\caption{Schematic view of the LUX detector. Figure adapted from \protect\cite{LUX_2013NIM}.} \label{LUX_scematic}
\end{center}
\end{figure}

LUX is a two-phase (liquid/gas)  time-projection chamber (LXe-TPC) containing 250 kg of xenon in the active volume of the detector\cite{LUX_2013NIM}. The schematic view of the LUX detector is shown on the figure \cref{LUX_scematic}. The principle of working and advantages of a LXe-TPC, common to other detectors such as ZEPLIN-III\cite{Akimov200746}, PandaX\cite{PhysRevLett.117.121303}, and XENON1T\cite{1475-7516-2016-04-027}, have been thoroughly studied and discussed in the literature\cite{RevModPhys.82.2053, Chepel2013}. In these detectors, a particle interacting in the liquid volume produces primary scintillation (called \emph{S1} signal) and charge through the ionization of the xenon atoms. An electric field, applied in the liquid phase, drifts the electrons towards the surface. These electrons are then extracted to the gas phase where they produce electro-luminescence (called \emph{S2} signal). Both signals ($\lambda$=178~nm) are detected by photomultipliers (122 PMTs in the LUX detector\cite{LUX2013_PMTs}) usually set in two symmetric arrays placed above and below the sensitive volume. The z-position of the event along the drift field is obtained by the difference in the arrival time between the \emph{S2} and the \emph{S1} signal while the $(x, y)$ position is obtained in a pattern analysis of the \emph{S2} light distribution in the top array\cite{Solovov2011_PositionReconstruction}. 

The LUX TPC has a dodecagonal shape with an internal maximum diameter of 48~cm. The drift field is defined between the cathode located 5~cm above the PMT windows of the bottom array and the gate grid located 48~cm above the cathode. The surface of the liquid xenon is located $\sim$5~mm above the gate. The anode, which sets both the extraction and electro-luminescence fields, is located 5~mm above the liquid level and 5~cm below the PMT windows of the top PMTs. 


To ensure a high light collection for both \emph{S1} and \emph{S2}, polytetrafluoroethylene (PTFE, also known as Teflon\textsuperscript{\textregistered{}}) reflector covers the inner surfaces of the TPC (reflectance >97\% in the liquid phase\cite{FranciscoNeves2017_Reflectancia}). The sensitive volume is surrounded by two low radioactive titanium cryogenic vessels \cite{LUXTitanium}. The cryostat is immersed in cylindrical ultra-pure water tank with 7.6~m width and 6.1~m diameter. The water tank protects the detector from gamma-rays and neutrons caused by radioactivity in rock and cosmic-ray muons. The water tank is instrumented with 20 PMTs in order to detect Cherenkov light produced by cosmic-ray muons and their secondaries.

The signals from the photomultipliers are processed and digitized using a custom-built data acquisition (DAQ) system\cite{LUXDAQ2012}. In this DAQ system, the signals are initially shaped with a two-stage amplifiers before being recorded in a dedicated 16 Struck 8 channel ADC modules with a sampling of 10 ns. The DAQ threshold is set such that 95\% of all single photoelectron (phe) pulses in each PMT are recorded.

PMT signals are calibrated in units of detected photons (phd) instead of photoelectrons (phe) to account for the double photon emission by single photons at the photocathode\cite{Faham2015}. For \emph{S1}<100~phd,  photon counting\cite{Savage2009} is used instead of the pulse area to estimate the size of the \emph{S1} pulse. This method is more advantageous when compared the pulse area method because it avoids the degradation in the resolution caused by the width of the single-detected-photon area distribution. Photon counting is, however, limited to low photon fluxes due the pile-up caused by the merging of two photons within a single count. In LUX, the pile-up of the \emph{S1} pulse is caused by the merging of two photons within a single count is corrected using a Monte-Carlo simulation. 

\section{LUX Data Analysis}

\subsection{Position corrections}
The detection efficiency of both scintillation (\emph{S1}) and ionization (\emph{S2}) signals  depends on the position of emission due to differences in the light collection and the presence of electronegative impurities. To account for these effects,  LUX detector is calibrated weekly with a $^{\mathrm{83m}}$Kr internal source\cite{Kr83PhysRev, hertel2015}. This metastable isomer is injected into the xenon gas circulation system and allowed to spread uniformly throughout the active volume. In the LUX analysis, $^{\mathrm{83m}}$Kr has been considered as mono-energetic source with a total energy deposition of 41.5~keV, providing a standard candle that can be used to correct both signals for differences in the light collection\footnote{In the WS2014-2016 data, position corrections are more complex that the picture presented here due the presence of a strong radial field which influences light and charge yield. Details can be found in \cite{LUX2016_SSR}.}. Here, \emph{S1} and \emph{S2} variables correspond to the corrected pulse size normalized to the center and top of the detector respectively while \emph{S1}$_{\mathrm{raw}}$ and \emph{S2}$_{\mathrm{raw}}$ correspond to the uncorrected quantities.

$^{\mathrm{83m}}$Kr calibrations are also essential for the position reconstruction as a necessary step to generate a set of empirical light response functions. These functions are used to predict the response of the PMTs as a function of the position of the emission of the \emph{S2} light\cite{Solovov2011_PositionReconstruction, PR2017}. These predictions are compared with the the observed response of the PMTs using a maximum likelihood  test to get the best position of the interaction that describes the observed results.

The reconstructed positions using the \emph{S2} light signal only match the original position of interaction in the liquid volume when the electric field is perpendicular to the liquid surface for all the positions within the active volume. However, we observed that  this is not strictly true. On the WS2013 data, we observed the presence of a small constant radial component in the electric field that slightly moves the drifted electrons  radially inwards during their upward drift (radial drift velocity of $\sim$0.1~mm/$\mu$s in the centripetal direction). A simulation of the electric field in the LUX detector with COMSOL Multiphysics\cite{COMSOL} showed that this effect was consistent with the transparency of the cathode and gate grids\cite{Fields2017}, causing a leakage field from below the cathode. Between WS2013 and WS2014--16, the detector has been under tests that aimed at increasing the voltage applied to the field grids. During these tests, the voltage of the grids was increased for an extended period of time above the onset of discharge in order to burn and remove protruded tips present in the grids. In WS2014--16 data, after the grid tests\cite{LUX2015_ReanalysisPRL}, the observed intensity of the radial field was stronger (radial drift velocity of $\sim$0.5~mm/$\mu$s), and it was observed to increase in magnitude over the course of the exposure. COMSOL simulations showed that this was consistent with a nonuniform and time-varying negative charge density in the PTFE panels that surround the liquid xenon sensitive volume (charge density of -3.6~$\mu$C/m${}^2$ at the start of the exposure increasing to -5.5~$\mu$C/m${}^2$ at  end of the run). This charge is believed to be a collateral result of the PTFE exposure to coronal discharge during grid tests.

Due the non-uniformity of the field, we need to distinguish the observed coordinates of the \emph{S2} light emission ($r_{\mathit{S2}}$ for the radius and  $\phi_{\mathit{S2}}$ for the azimuth) from the true position of the energy deposition ($r_{\mathrm{ver}}$ for the radius and  $\phi_{\mathrm{ver}}$ for the azimuth). In WS2013 data analysis, the reconstructed positions from the \emph{S2} light emission ($r_{\mathit{S2}}$, $\phi_{\mathit{S2}}$) were corrected to the true position of the energy deposition. These corrections were obtained  from a look-up-table that was created assuming that the $^{\mathrm{83m}}$Kr events are distributed uniformly in the chamber (more details in\cite{hertel2015}). In WS2013, the depth of the event, $z$, is related to the drift time $\tau_d$ of the event by
\begin{equation}
z = v_d\cdot\tau_d = (0.1518\pm0.0011)\tau_d, \quad r_{\mathrm{ver}}<20~\mathrm{cm}
\end{equation}
where $v_d$ is the average drift velocity. In this case, the cathode position for $r_{\mathrm{ver}}$<20~cm corresponds to a drift time of 322.3$\pm$0.4~$\mu$s.

In  WS2014--16, the event position is determined and stored in observed coordinates ($r_{\mathit{S2}}, \phi_{\mathit{S2}}$) and the drift time $\tau_d$, while the true positions, given by the simulations, are mapped into the observed positions. The depth $z$ cannot be approached to a linear dependence as before because  $v_d$ depends both on the position and the absolute time of the event due to the non-uniformity of the field and its dependence on date resulting in the radial drift velocity to increase with time. In this case, the cathode position, as converted to drift time, is dependent on these factors and varies between 320~$\mu$s and 400~$\mu$s. The mapping used to relate the real positions $(r_{\mathrm{ver}}, \phi_{\mathrm{ver}}, z)$ with the observed positions $(r_{\mathit{S2}}, \phi_{\mathit{S2}}, \tau_d)$ was created by the COMSOL simulations.


\subsection{ER and NR calibrations}
Most of the background sources (such as $\gamma$ and $\beta$ decays) produce electron recoil events (ERs) while the signal from WIMPs, or neutrons, scatter off the target nucleus producing nuclear recoil events (NRs). The \emph{S2}/\emph{S1} ratio is the critical ER/NR discriminating parameter used on an event-by-event basis. The \emph{S2} is smaller in an NR event, when compared with an ER recoil for the same value of \emph{S1} due to a higher recombination rate in a NR event.

The LUX experiment has developed new methods for calibrating the detector response to ERs and NRs. The detector response to low-energy ERs is obtained with a tritiated methane (CH$_{3}$T) $\beta^{-}$ internal source\cite{TritiumPaper2015} (end-point energy 18.6 keV). 2.45~MeV neutrons emitted by a Deuterium-Deuterium (D-D) generator are used to characterize the detector response to NRs\cite{LUX2016_DDCalibrations}. These neutrons are  collimated using an air-filled tube going through the water tank to the walls of the detector. In such configuration, the deposited energy can be determined directly from the scattering angle in double scatter events. Both D-D and CH$_{3}$ calibrations were performed after the end of WS2013 run between October and December 2013 and about every three months during WS2014--16 run.


Interpretation of ER and NR calibrations is complicated in WS2014--16 data due  to the large variation in the magnitude of the electric field in the active region (500-650~V/cm near the top to 50-20~V/cm near the bottom \cite{Fields2017}). The dependence of the charge and light yields on the electric field is taken into account by dividing the exposure in 4 different time periods  with each period further divided into four equal bins in drift time. The start and end dates of the time periods are: September 11, 2014; January 1, 2015; April 1, 2015; October 1, 2015; May 2, 2016. The boundaries in the drift time are: 40, 105, 170, 235, 300~$\mu$s. Each exposure bin has unique  response model determined by the calibrations. The response model of each bin is obtained using the Noble Element Simulation Technique (NEST)\cite{NEST2011} code which provides an underlying physics model based only on detector parameters. 

The reconstructed energy of ER events is obtained through a combination of the \emph{S2} and \emph{S1} signals\cite{TritiumPaper2015}:
\begin{equation}
\label{eq:etotal}
E({\rm keV_{ee}}) = W\cdot(n_e + n_{\gamma}) = W\cdot\left(\frac{\mathit{S1}}{g_1} + \frac{\mathit{S2}}{g_2}\right) = W\cdot\left(\frac{\mathit{S1}}{g_1} + \frac{\mathit{S2}}{SE\cdot EE}\right)
\end{equation}
The gains $g_1$ and $g_2$ convert \emph{S1} and \emph{S2}  to number of photons ($n_{\gamma}$) and electrons ($n_e$), respectively. $EE$ is the extraction efficiency at the liquid gas interface and $SE$ is the average response in the number of detected photons phd from a single extracted electron. $W$ is a constant assumed to be $W$ = 13.7$\pm$0.2~eV \cite{Dahl_2009}. For both runs, the values of $g_1$, $g_2$, are measured  using a set of mono-energetic electronic recoil sources \cite{Doke2002}. The values of $EE$, $g_1$, $g_2$ for both runs are shown in Table 1. 


From December 8th 2014 until the end of the exposure, a number of artificial WIMP-like events have been added to the data pipeline (this procedure has been called 'salting') and thus blinding the potential WIMP signal. The \emph{S1} and \emph{S2} pulses that went into the artificial events were taken from tritium calibrations and combined in such a way so they looked like NR events. These artificial events have been uniformly distributed in the detector volume and time of acquisition and could not be distinguished from a real detector signal. The trigger time of each salt event is kept by a person outside the LUX collaboration, and it cannot be accessed by any LUX collaboration member until the main analysis parameters such as quality cuts, efficiencies, and background models have been established.

\subsection{Quality cuts}

To remove most of the events originated by background radiation or noise, we applied a sequence of quality cuts. Some of these cuts are applied to both WS2013 and WS2014--2016. Those are:
\vspace{-0.4cm}
\paragraph{Single Scatter Cut:} We select events with only a single valid \emph{S1} pulse before an \emph{S2} and a single valid \emph{S2} after the respective \emph{S1}. The \emph{S2}$_{\mathrm{raw}}$ must be larger than 55~phd. The lowest \emph{S1} signal is given by the coincidence of two photon signals in two different PMTs. 
\vspace{-0.4cm}
\paragraph{Detector Stability Cut:} Periods of live-time when the detector was not stable are excluded. To determine these periods, we monitored the stability of several slow control parameters (e.g.\ liquid level, HV grid voltages and currents, etc.) that would affect the response of the detector. When the value of any of those sensors was outside the predefined range of normal operation, the data was excluded.
\vspace{-0.4cm}
\paragraph{High Electron Life-Time Cut:} Periods in which the electron life-time was lower than 500 $\mu$s were excluded. 
\vspace{-0.4cm}
\paragraph{Low Rate of Single Electron Background Cut:} Events with a high rate of single electron background were excluded. In these events, large \emph{S2} signals are followed by an extended tail of single electron emitted from the liquid phase that might last several milliseconds\cite{LUX2015_ReanalysisPRD}. Some of these single electrons can be grouped together exceeding the \emph{S2} threshold and considered a valid  \emph{S2}. They may be reconstructed by the pulse finding algorithm to an additional isolated \emph{S1} signal. These events are identified by their large fraction of the total pulse size of the event outside the \emph{S1} and \emph{S2} signals.
\vspace{-0.4cm}
\paragraph{Fiducial Volume:} 
A very important group of background sources are created from ${}^{210}$Pb  subchain isotopes  platted on the PTFE walls or field grids. Those events cannot be discriminated based on the ratio \emph{S2}/\emph{S1}  because a fraction of the extracted charge is lost in the PTFE walls. Additionally, a low energy NR recoil can be observed as a result of the ${}^{210}$Po decay. This source emits a 5.3~MeV  $\alpha$-particle that can be absorbed by the PTFE walls. In this case the only signal observed corresponds to the nuclear recoil of the ${}^{206}$Pb nucleus. Most of the events from ${}^{210}$Pb  subchain  can be removed by applying a fiducial cut, i.e. looking for signals only in the inner region of the detector. This cut removes also a significant fraction of the background events originated from the $\gamma$-- and $\beta$--ray sources. The fiducial cuts that were applied in each run are shown in Table 1. 
\vspace{-0.4cm}
\paragraph{\emph{S2}$_{\mathrm{raw}}$ threshold:} The threshold of \emph{S2}$_{\mathrm{raw}}$ is set to 165~phd for the WS2013 and to 200~phd for the WS2014--16. This removes events leaking from the walls due to a high uncertainty in the position reconstruction (the position uncertainties are proportional to 1/$\sqrt{\mathit{S2}_{\mathrm{raw}}}$).

Additionally to these cuts, the range of both \emph{S1} and \emph{S2} where the WIMPs are expected to be is shown on the Figure 1.
\vspace{0.4cm}

The salting in the data allowed to impose additional  quality cuts  to the WS2014--16 data. These cuts were developped using calibration data (both tritium and D--D data).
\vspace{-0.4cm}

\paragraph{\emph{S2} Pulse Width Cut:} $\sigma_{S2}$ is the standard deviation  that results from a Gaussian fit to the \emph{S2} pulse waveform. This cut is set to  $\sigma_{S2}$>0.4\,$\mu$s. It removes events with energy depositions in the gas region below the anode. In these events,  the \emph{S1} and \emph{S2} are merged and the Gaussian fit is often performed to the \emph{S1} or to a short spike in the \emph{S2}.
\vspace{-0.4cm}
\paragraph{\emph{S2} Pulse Quality Cut:} We imposed an upper and lower limit on the ratio  $t_{50\%a}/\sigma_{S2}$  where $t_{50\%a}$ corresponds to the time difference between the cumulative 1\% and 50\% area fraction of the \emph{S2}. This cut tags \emph{S2} pulses generated from multiple-scatter events that are emitted with a very short time difference for the \emph{S2}, being merged into a single \emph{S2} pulse. 
\vspace{-0.4cm}
\paragraph{Position Reconstruction $\chi^2$ Cut:} Events from multiple-scatters separated on $(x, y)$ that are merged into a single \emph{S2} pulse, or events with a PMT affected by after-pulsing have a distribution of light among the PMTs clearly distinct from a \emph{good} \emph{S2} pulse. The minimum of the $\chi^2$ obtained from the position reconstruction method is sensitive to this, and it can be used to remove those pathologies. 
\vspace{-0.4cm}
\paragraph{\bf \emph{S2} cut: }
The value of \emph{S2} is  limited between $M_{NR}$-5$\sigma_{NR}$ and $M_{ER}$+3$\sigma_{ER}$, where $M$ corresponds to the band median and $\sigma$ to the standard deviation of the band. 
\vspace{-0.4cm}

\vspace{0.4cm}

The acceptance of the cuts, applied exclusively to the WS2014--16 data, was studied using a known population of single scatter ${}^{3}$H events. The combined acceptance for the cuts applied to the \emph{S2} pulse topology is 65\% at \emph{S2}=200~phd rising to $95\%$ above 1000~phd.

After the salt was removed from the data, we identified two small populations of events with \emph{S1} pulse topology that could not arise from energy depositions in the liquid. For this reason, we applied two additional post-salting cuts:

\vspace{-0.4cm}

\paragraph{\bf \emph{S1} Pulse Fraction Cut:} In a very low number of events, a significant fraction of the \emph{S1} pulse is observed in a single PMT.  Those pulses are caused by after-pulsing in a single PMT or other light source outside the TPC. They were identified by imposing a maximum waveform area in an individual PMT as function of \emph{S1}$_{\mathrm{raw}}$ (summed over all PMTs).
\vspace{-0.4cm}
\paragraph{\bf \emph{S1} Prompt Fraction Cut:} The prompt fraction corresponds to the fraction of the \emph{S1} pulse area within the initial 120~ns of the pulse. This cut removes energy depositions in the gas phase, characterized by a long decay constants on the \emph{S1} pulse.

These two cuts were tuned with calibration data and have a very high acceptance of more than that 99\% across all energies.

\begin{figure}
\begin{center}
\includegraphics[width=\textwidth]{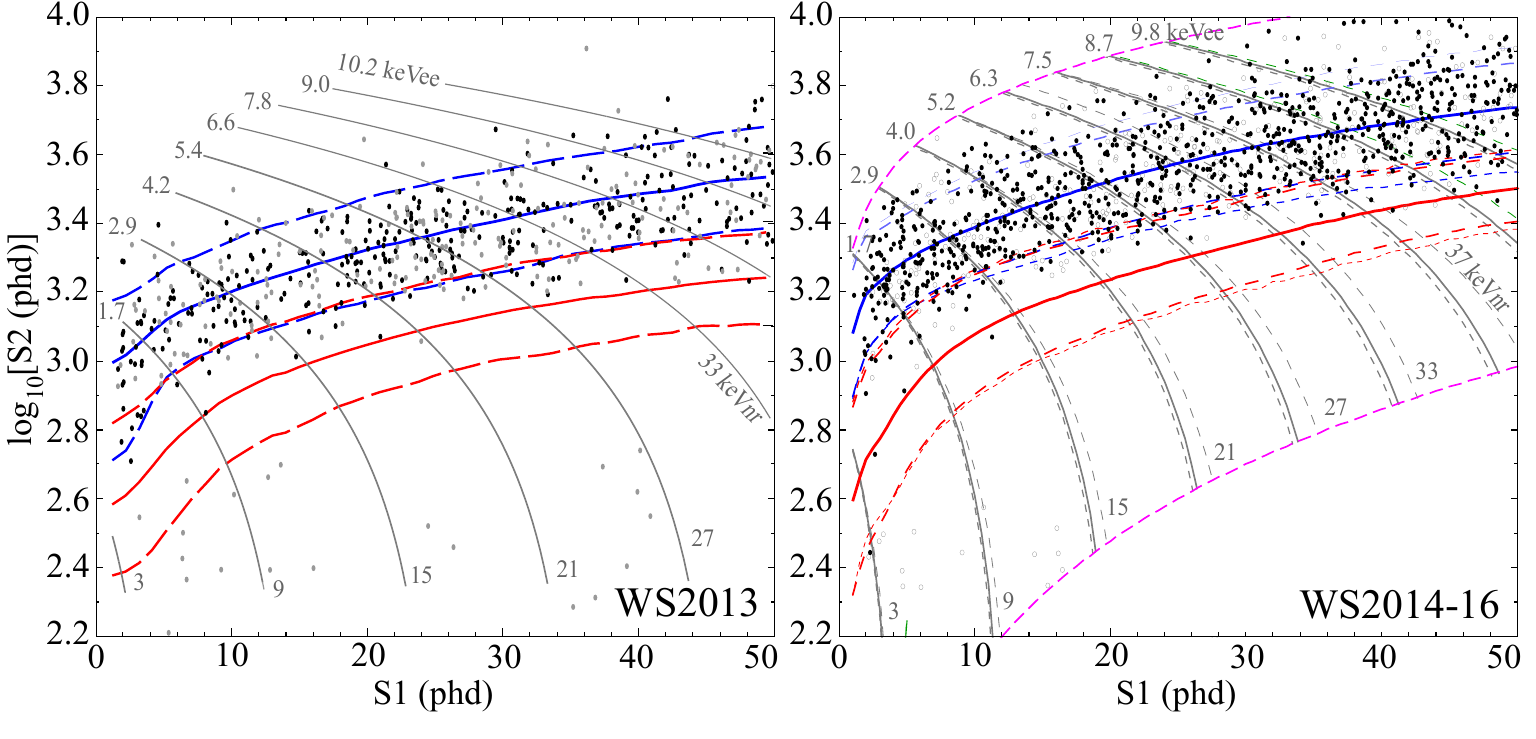}
\caption{WIMP search data passing the quality cuts for the WS2013 data (left panel) and WS2014--16 data (right panel). For the WS2013, gray dots indicate events with a reconstructed radius between 18 and 20~cm. For the WS2014--2016, unfilled circles indicate events within 1 cm of the radial fiducial volume boundary (distance to the wall between 4 and 3~cm). ER and NR bands are indicated in blue and red, respectively. The dashed lines correspond to the 10\% and 90\% contours. Gray curves indicate mean energy contours for ER event (top labels)
and for NR events (lower labels). For the WS2014--16, the red and blue curves correspond to an exposure-weighted average with the fainter dashed lines representing the scale of variation over the 16 data bins.} \label{LUX_Data}
\end{center}
\end{figure}

The data passing all the selection criteria is presented on \cref{LUX_Data} for the WS2013 data (left panel) and WS2014--16 data (right panel). The events with a low value of \emph{S2} and closer to the walls (identified on WS2013 by the gray circles and WS2014--16 by the unfilled circles) are most probably events from ${}^{210}$Pb  subchain that loose charge to the PTFE.

\begin{table}
\label{GreatTable}
\begin{center}
\begin{threeparttable}
{\caption{Main detector and analysis parameters used in runs WS2013 and WS2014--16.}}
\begin{tabular}{lcc}
\hline 
                                    & WS2013           & WS2014--16                          \\

Live-days                           & 95               & 332                                 \\
Total Exposure (kg$\cdot$live-years) & 38.3 & 95.8\\
Anode Voltage (kV) & +3.5 & +7.0\\
Gate Voltage (kV)  & -1.5 & +1.0\\
Cathode Voltage (kV) & -10.0 & -8.5\\
$EE$ - Extraction Efficiency (\%)      & 49$\pm$3         & 73$\pm $4                        \\
$g_1$  (phd per photon)             & 0.117$\pm$0.003  & 0.100$\pm$0.002 to 0.097$\pm$0.001  \\
$g_2$  (phd per extracted electron)    & 12.1$\pm$0.8     & 18.9$\pm$0.8 to 19.7$\pm$2.4        \\
Radial cut\textsuperscript{\dag}   &$r$<20~cm &$d_{\mathrm wall}$>3~cm\\
Drift time cut ($\mu$s)& 38-305 & 40-300 \\
Fiducial Mass  (kg)       &145.4$\pm$1.3             & 105.4-98.4\\
\emph{S1} range (phd)\textsuperscript{$\ast$}   & 1 - 50      & 1 - 50  \\
\emph{S2} (phd, $M$ is band median)                    &        <10\,000         & <10\,000 and \\&&$M_{NR}$-5$\sigma_{NR}$<\emph{S2}<$M_{ER}$+3$\sigma_{ER}$\\
\emph{S2}$_{\mathrm{raw}}$ cut (phd) & >165          & >200\\
Number of events after cuts & 591 & 1\,221 \\
\hline
\end{tabular}
\begin{tablenotes}
\item[\dag] $r$ corresponds to the radius measured in corrected variables while $d_{\mathrm wall}$ is the distance to the wall measured in uncorrected variables (\emph{S2} position of light emission).
\item[$\ast$] 2 PMTs with at least 1 detected photon is required.
\end{tablenotes}
\end{threeparttable}
\end{center}
\end{table}

\section{Likelihood Analysis}

A Profile Likelihood Ratio (PLR) analysis was used to compare the null (background-only) hypothesis and signal plus background hypothesis. The results presented here are a combination of WS2013 data and WS2014--16 data on event by event basis and not by a simple combination of the limits. To achieve this, we treated the data from WS2013 as the 17th exposure segment with the detector parameters and analysis cuts as specified in Table 1. The likelihood for both the signal and background is modeled as a function of the variables \{$r_{\mathrm{ver}}$, $z_{\mathrm{ver}}$, \emph{S1} , \emph{S2}\} for the WS2013 and \{$r_{\mathit{S2}}$, $\phi_{\mathit{S2}}$, $\tau_d$, \emph{S1} , \emph{S2}\} for each one of the 16 bins of WS2014-2016 data\cite{LUX2016_SSR} ($\tau_d$ corresponds to the drift time).

The description of the background model in the PLR includes three different types of events: i) events with a typical charge and light yield of an ER ($\gamma$s and $\beta$s) and NR (${}^{8}$B solar neutrinos); ii) wall events from ${}^{210}$Pb  subchain that are affected by charge suppression or are of the NR type such as the signal produced from the recoiling of ${}^{206}$Pb; and iii) random coincidences between an \emph{S1} and an \emph{S2} pulse. The details of the background model are described in WS2013 and WS2014--16 articles \cite{LUX2015_ReanalysisPRL, LUX2016_SSR} and in a  background dedicated paper\cite{LUX2015_BackgroundPaper}. In the WS2013 run, the \emph{p}-value obtained using the Kolmogorov-Smirnov test was larger than 0.05 for the probability distribution function projections for each one of the 4 observables. For the WS2014--16 run, the \emph{p}-value obtained using the same test was larger than 0.6 for the PDF projections for each one of the 5 observables. This shows a good agreement between the data and the background model.
\begin{figure}[h]
\begin{center}
\includegraphics[width=0.45\textwidth]{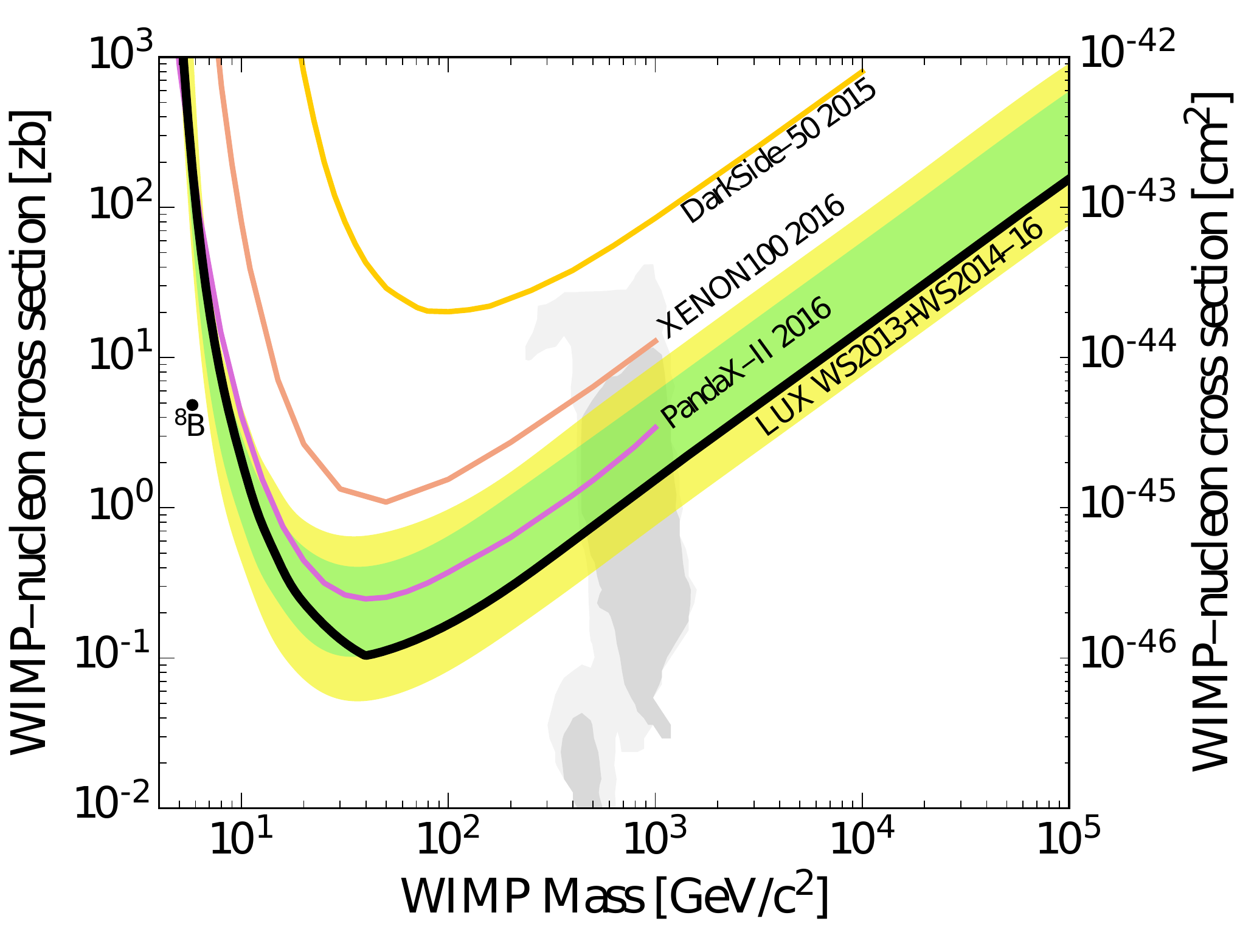}

\includegraphics[width=0.44\textwidth]{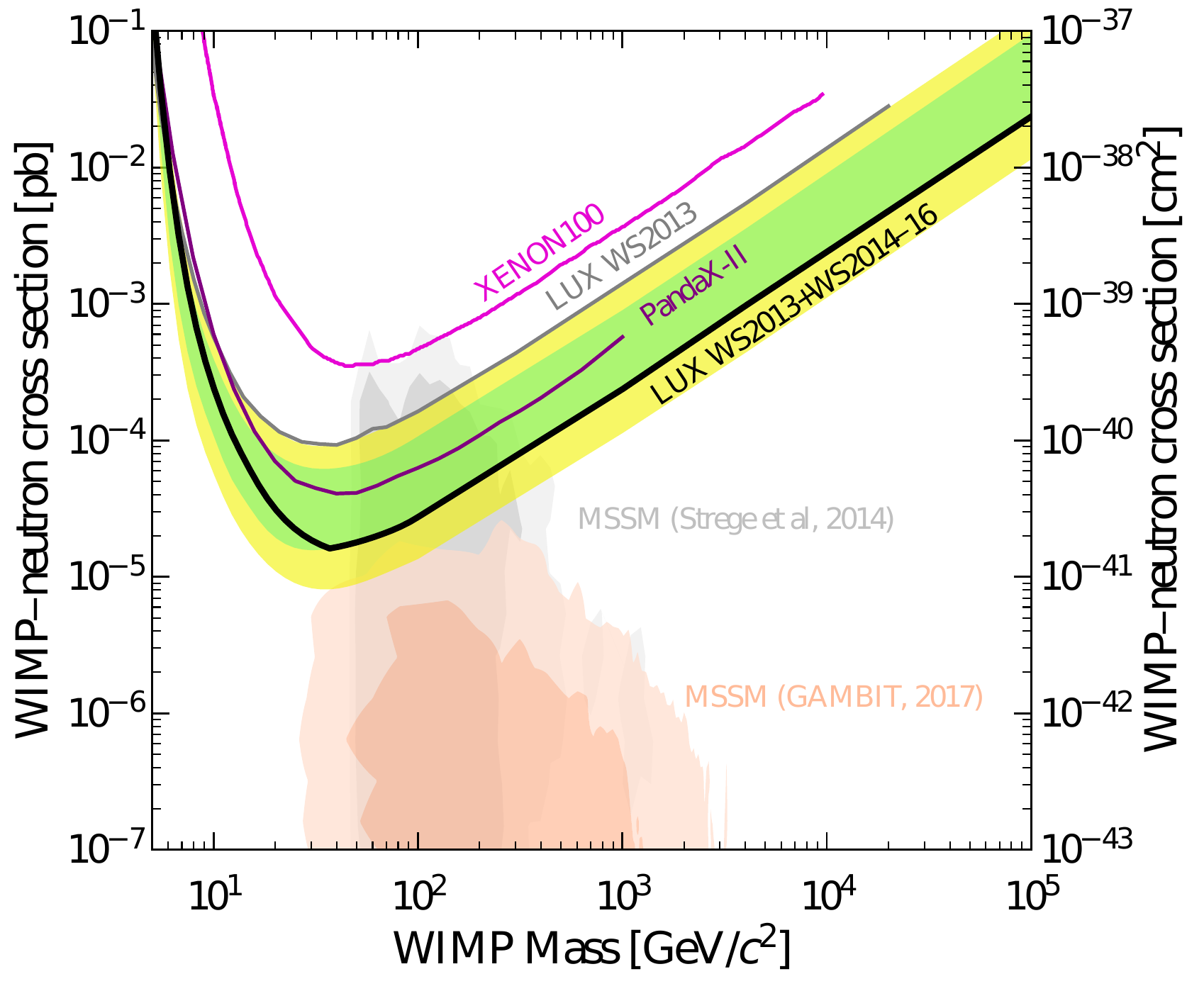}
\includegraphics[width=0.44\textwidth]{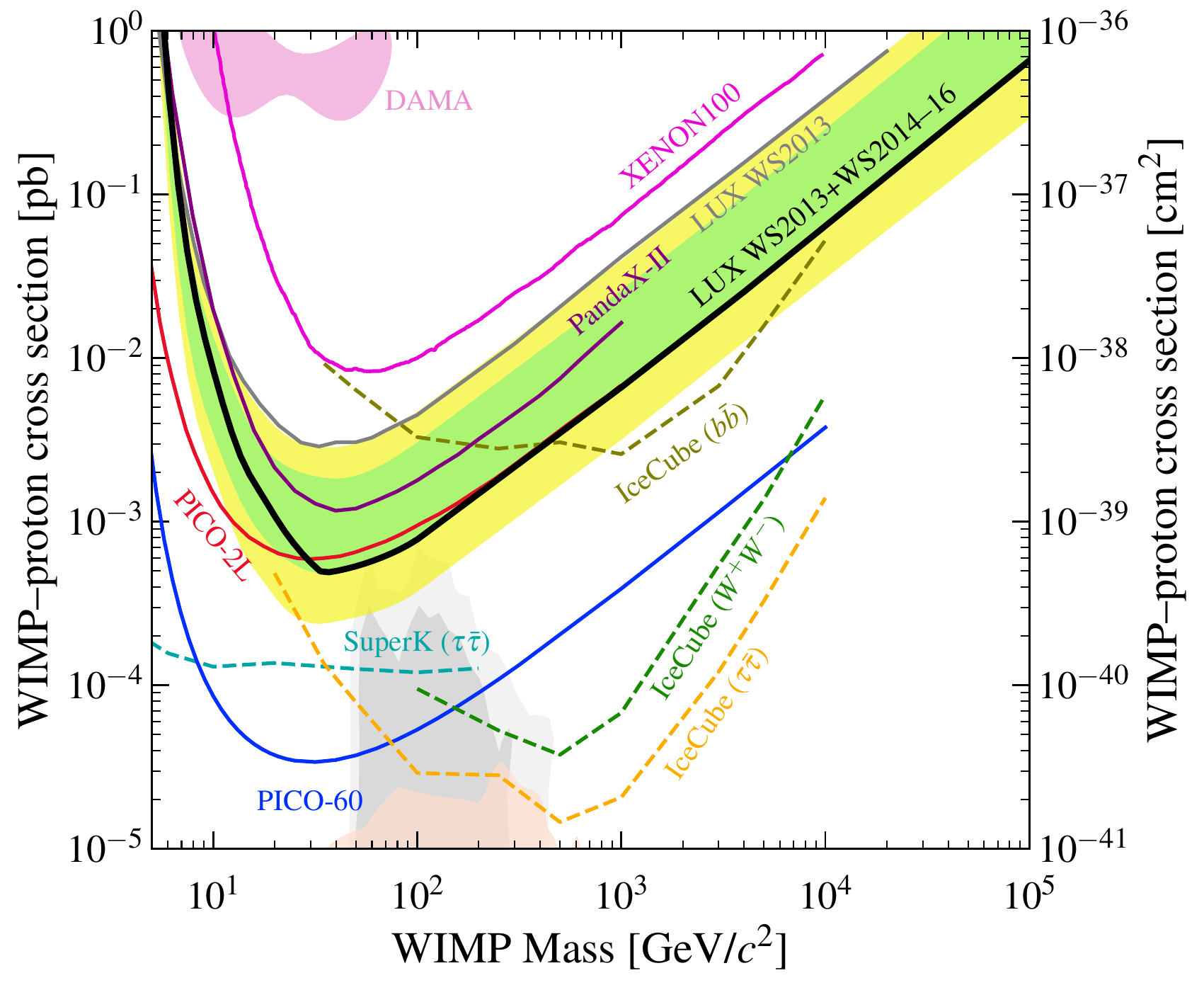}
\caption{Spin-independent WIMP-nucleon (top panel) and spin-dependent WIMP-neutron/WIMP-proton (bottom panels) exclusion limits at 90\% C.L. for the combined results (WS2013+ WS2014--16) are shown by the black lines.  The 1-$\sigma$ and 2-$\sigma$ ranges of background-only trials for this combined result are shown as green and yellow bands, respectively. Constraints from other LXe TPC experiments are also shown, including XENON100\protect\cite{XENONSD} and PandaX-II\protect\cite{PANDA_SD}. On the spin-independent results, the parameters favored by SUSY CMSSM\protect\cite{Bagnaschi2015} before this result are indicated as dark and light gray (1-$\sigma$ and 2-$\sigma$) filled regions. On the spin-dependent interaction, the gray regions corresponds to the profile likelihood maps obtained via global fits of a phenomenological Minimal Supersymmetric Standard Model with 15 free parameters (MSSM15) obtained by \cite{Strege2014}. The results from the GAMBIT collaboration using a seven-dimensional Minimal Supersymmetric Standard Model (MSSM7) are represented by the red region  \cite{GAMBIT2017}.}
\label{FIG_Run3andRun4LimitPlot}
\end{center}
\end{figure}

The recoil spectra for the WIMP signal are described in \cite{LUX2016_SSR, LUX2017_SDCombined}, assuming the same WIMP halo model as  in other experiments such as XENON\cite{XENON1T2017, XENON100_477} and CDMS\cite{CDMS_Results} with an average earth velocity of 245~km/s during the WS2013 data taking and 230~km/s during WS2014-16\footnote{For the WS2014-16 spin-independent limits, we considered an average earth velocity of 245~km/s.}. The elastic scattering of the WIMP with the nucleus may be generally described as arising from spin-independent (SI) and spin-dependent (SD) WIMP-nucleon interactions \cite{EFT2013, Undagoitia2016}. Confidence limits on a given type of cross section (SI or SD) are calculated by assuming it to be the sole coupling mechanism. SI coupling implies coherent scattering with nucleons in the nucleus, so heavier target isotopes receive an $A^2$ enhancement (proportional to the atomic mass of the nucleus). SD WIMP-nucleus scattering does not benefit from coherent enhancement, and depends on the spin structure factor of the isotope. As a result, since experiments are typically more sensitive to either WIMP-proton or WIMP-neutron SD coupling, these limits are calculated separately \cite{XENONSD, LUXSD2016, LUX2017_SDCombined}. We also computed exclusion limits for different WIMP masses on the more general parameter space of WIMP-proton and WIMP-neutron coupling constants using the procedure detailed in  \cite{Tovey200017}.


The combined WS2013+WS2014--16 exclusion limits for both spin-independent WIMP-nucleon and spin-dependent WIMP-neutron/WIMP-proton cross sections at 90\% confidence level are shown in \cref{FIG_Run3andRun4LimitPlot}.  The exclusion limit reaches a  minimum of 0.11$\times$10$^{-45}$~cm$^2$ for the spin-independent interaction  at  50 GeV/c$^2$. For the spin-dependent interaction, LUX reaches a minimum  of 1.6$\times$10$^{-41}$~cm$^2$ for the neutron-only coupling and  of 5.0$\times$10$^{-40}$~cm$^2$ for the proton-only coupling (both at 35 GeV/c$^2$).

\section{Conclusion and perspectives}

In the  four years of operations LUX achieved the world leading result in sensitivity for both spin-independent and spin-dependent (WIMP-neutron coupling) cross section. No signal due to a possible WIMP particle was identified and a significant fraction of the WIMP parameter space was excluded. 

Major advances in the detector calibration have been reported: internal source of tritiated methane and 2.45 MeV neutrons from a D-D generator have been used to determine the detector response to ERs and NRs, respectively. 

The analysis of  LUX data will continue through the year of 2017 with a search for other possible  signals. Results for the first searches for axions and axion-like particles were already presented\cite{LUXAxions2017}. 

In 2019, a new very massive detector called LUX-ZEPLIN (LZ) will be installed in the same location as the LUX detector\cite{LZ2017}. This new detector will feature 7 tonnes of xenon in the active region, and will be able to improve the current sensitivity to the spin-independent WIMP-nucleon cross section by a factor of 50 during 1\,000 live-days of operations. LZ will be operational in 2020.

\section*{Acknowledgments}

{\footnotesize {\linespread{0.85}\selectfont 
This work was partially supported by the U.S. Department of Energy under Awards No. DE-AC02-05CH11231, DE-AC05-06OR23100, DE-AC52-07NA27344, DE-FG01-91ER40618, DE-FG02-08ER41549, DE-FG02-11ER41738, DE-FG02-91ER40674, DE-FG02-91ER40688, DE-FG02-95ER40917, DE-NA0000979, DE-SC0006605, DE- SC0010010, DE-SC0015535, and DE-SC0014223, the U.S. National Science Foundation under Grants No. PHY-0750671, PHY-0801536,  PHY-1003660,  PHY-1004661, PHY-1102470,  PHY-1312561,  PHY-1347449,  PHY-1505868, and  PHY-1636738, the Research Corporation Grant No. RA0350, the Center for Ultra-low Background Experiments in the Dakotas, and the South Dakota School of Mines and Technology. LIP-Coimbra acknowledges funding from Funda\c c\~ao para a Ci\^encia e a Tecnologia through the Project-Grant No. PTDC/FIS-NUC/1525/2014. Imperial College and Brown University thank the UK Royal Society for travel funds under the International Exchange Scheme (Grant No. IE120804). The UK groups acknowledge institutional support from Imperial College London, University College London and Edinburgh University, and from the Science \& Technology Facilities Council for Ph.D. studentships Grants No. ST/K502042/1 (A. B.), ST/ K502406/1 (S. S.), and ST/M503538/1 (K. Y.). The University of Edinburgh is a charitable body registered in Scotland, with Registration No. SC005336. 
We gratefully acknowledge the logistical and technical support and the access to laboratory infrastructure provided to us by SURF and its personnel at Lead, South Dakota. SURF was developed by the South Dakota Science and Technology Authority, with an important philanthropic donation from T. Denny Sanford, and is operated by Lawrence Berkeley National Laboratory for the Department of Energy, Office of High Energy Physics.
 \par}}

\end{document}